\documentclass[cits]{PoS}
\pdfoutput=1
\title{Accelerating Twisted Mass LQCD with QPhiX}

\ShortTitle{Accelerating Twisted Mass LQCD with QPhiX}

\author{\speaker{Mario Schr\"ock}\\
       INFN - Sezione Roma Tre, Rome (Italy)\\
       E-mail: \email{mario.schroeck@roma3.infn.it}}

\author{Silvano Simula\\
       INFN - Sezione Roma Tre, Rome (Italy)\\
       E-mail: \email{silvano.simula@roma3.infn.it}}

\author{Alexei Strelchenko\\
        Fermi National Accelerator Laboratory\\
        E-mail: \email{astrel@fnal.gov}}        

\abstract{We present the implementation of twisted mass fermion operators for the QPhiX library\footnote{\codefont{https://github.com/JeffersonLab/qphix}}. We analyze the performance on the Intel Xeon Phi (Knights Corner) coprocessor as well as on Intel Xeon Haswell CPUs. In particular, we demonstrate that on the Xeon Phi 7120P the Dslash kernel is able to reach 80\% of the theoretical peak bandwidth, while on a Xeon Haswell E5-2630 CPU our generated code for the Dslash operator with AVX2 instructions outperforms the corresponding implementation in the tmLQCD library by a factor of $\sim 5\times$ in single precision. We strong scale the code up to 6.8 (14.1) Tflops in single (half) precision on 64 Xeon Haswell CPUs.}

\FullConference{The 33rd International Symposium on Lattice Field Theory\\
		 14--18 July  2015\\
		 Kobe International Conference Center, Kobe, Japan}

\usepackage[tight]{units}
\usepackage{amsmath}

\newcommand{\eq}[1]{(\ref{#1})}

\newcommand{\fig}[1]{Fig.~{\ref{#1}}}

\newcommand{\be}{\begin{equation}}
\newcommand{\ee}{\end{equation}}
\newcommand{\gaf}{\gamma_5}

\newcommand{\Dtm}{\not\!\!D_{\textrm{TM}}}
\newcommand{\Dw}{\not\!\!D_{\textrm{W}}}
\newcommand{\codefont}[1]{{\ttfamily #1}}

\begin{document}

\section{Introduction}

The \emph{QPhiX} library \cite{joo:2013} offers a collection of highly optimized kernels and inverters to perform Lattice QCD computations on recent Intel x86 systems.  
QPhiX is a C++ library and consists of two components: the lower level component of the library provides with the code generator which abstracts away vector intrinsics (supporting IMCI, AVX, AVX2, AVX512, SSE and QPX instructions). The high level part is concerned with parallelizing over threads via OpenMP and multi-processing via MPI; it is also performing the loop structure for the cache-blocking strategy.
Currently, the library supports Wilson (Clover) fermions and (mixed precicion) CG, BiCGstab inverters. There is also an implementation of the staggered Dslash operator \cite{Li:2014kxa}.
For details we refer to the original work \cite{joo:2013}.
In the present study we extend QPhiX for initial support of degenerate twisted mass fermions \cite{Frezzotti:2000nk, Frezzotti:2003xj}.

\section{QPhiX and twisted mass fermions}
The degenerate twisted mass Dslash operator $\Dtm$ is defined as
\be\label{eq:DslashTM}
	\Dtm = \Dw 1_{f} + i\mu\gaf\tau_3\,,
\ee
where $\Dw$ is the standard Wilson Dslash operator which is diagonal in flavour space.
The second term is the \emph{twisted mass term} which is nontrivial in flavour space ($\tau_3$ being the third Pauli matrix in flavour space), $\mu$ is the twisted mass.

QPhiX relies on even-odd preconditioning and therefore the operator \eq{eq:DslashTM} requires  two kernels (plus the corresponding daggered versions) which act on the even (odd) sites separately, in QPhiX parlance, \emph{Dslash} and \emph{AChiMBDPsi}:
\begin{enumerate}
	\item \emph{Dslash}:
	\be
		\chi = R^{-1} \Dw \psi
	\ee
	\item \emph{AChiMBDPsi}:
	\be
		\phi = R\chi - b\Dw\psi
	\ee
	with $R = 1 + i2\kappa\mu\gaf$ being the twist operator.
\end{enumerate}
The successive application of these kernels corresponds to the Schur decomposed fermion matrix
\be\label{eq:Moo}
	\tilde M_{oo} = R_{oo} - \frac{1}{2}\kappa (\Dw)_{oe} \,R^{-1}_{ee} \,(\Dw)_{eo}
\ee
which then enters, e.g., the Conjugate Gradient algorithm.
We implement these kernels on the level of the code-generator.
The high-level part has to be modified to include the twisted mass parameter $\mu$ and to call the corresponding twisted mass low-level kernels, apart from that it is equivalent to the pure Wilson case.

The QPhiX routine \codefont{dslash\_plain\_body()} generates the (Wilson) Dslash kernel routines,
within that, after the call of \codefont{dslash\_body()}, we call the (inverse) twisted term.
This generates files with vector intrinsics of the form, e.g.,
\begin{center}
\codefont{tmf\_dslash\_plus\_body\_float\_float\_v8\_s4\_12} 
\end{center}
for the different floating point precisions, SIMD vector lengths, structure of arrays lengths and gauge field compression types (12 or 18).
Next, the generated files are wrapped up as template specializations of the form
\begin{verbatim}
template<> 
inline void
dslash_plus_vec<FPTYPE,VEC,SOA,COMPRESS12>(...)
{
    #include INCLUDE_FILE_VAR(qphix/avx2/generated/dslash_plus_body_,
        FPTYPE,VEC,SOA,COMPRESS_SUFFIX)
}\end{verbatim}
The procedure for the AChiMBDPsi kernels is done in a close analogy with the plain Dslash ones.

Lastly, we have to consider the unpacking routines for the MPI communication. The spinors are projected to halfspinors before being exchanged at the boundaries of each domain. 
These packing routines are identical to the Wilson case. However, when unpacking, while accumulating the different directions and before streaming to memory, we still have to apply the twisted mass term and therefore we need twisted mass specific unpacking routines.

\section{Performance}
\subsection{Single node}

\begin{figure}[htb]
	\center
	\includegraphics[width=0.75\textwidth]{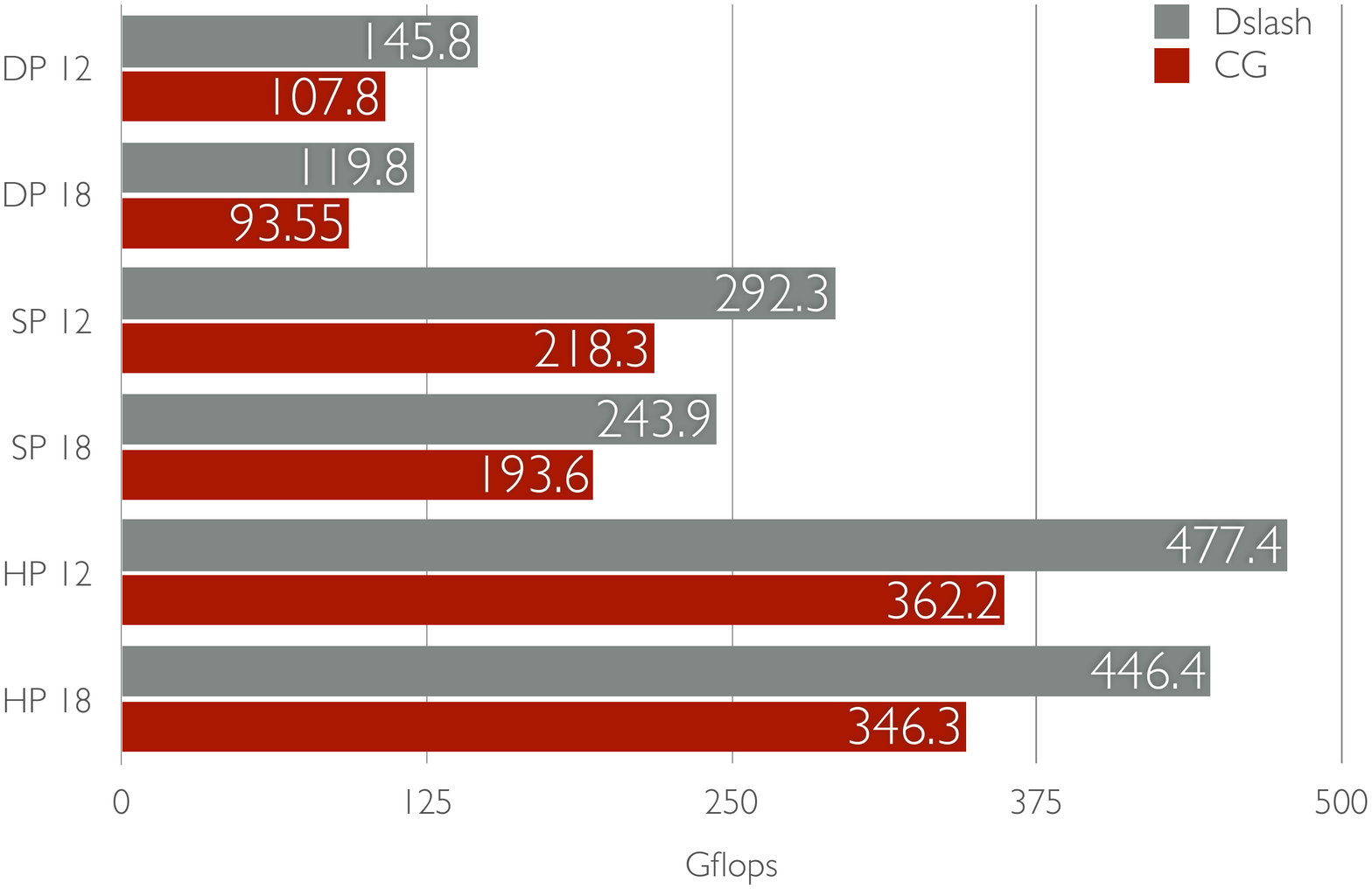}
	\caption{Twisted Mass $32^3\times 64$ Xeon-Phi 7120P.}
	\label{fig:tm_32x64_mic}
\end{figure}

In \fig{fig:tm_32x64_mic} we summarize the performance results of our twisted mass Dslash kernels as well as of the full Conjugate Gradient (CG) inversion algorithm on a single Intel Xeon-Phi 7120P. The results are distinguished by double (DP), single (SP) and half precision (HP) and moreover whether 12 compression\footnote{To overcome the bandwidth bottleneck to some extent, one may transfer only two of the three SU(3) matrix rows (thus 12 parameters instead of 18) and recalculate the third one on-the-fly.} has been used or not (otherwise 18 ``compression'').
Note that the DP and SP 12 compression Dslash kernels reach around 80\% of the theoretical peak bandwidth of the Xeon-Phi.

\begin{figure}[htb]
	\center
	\includegraphics[width=0.75\textwidth]{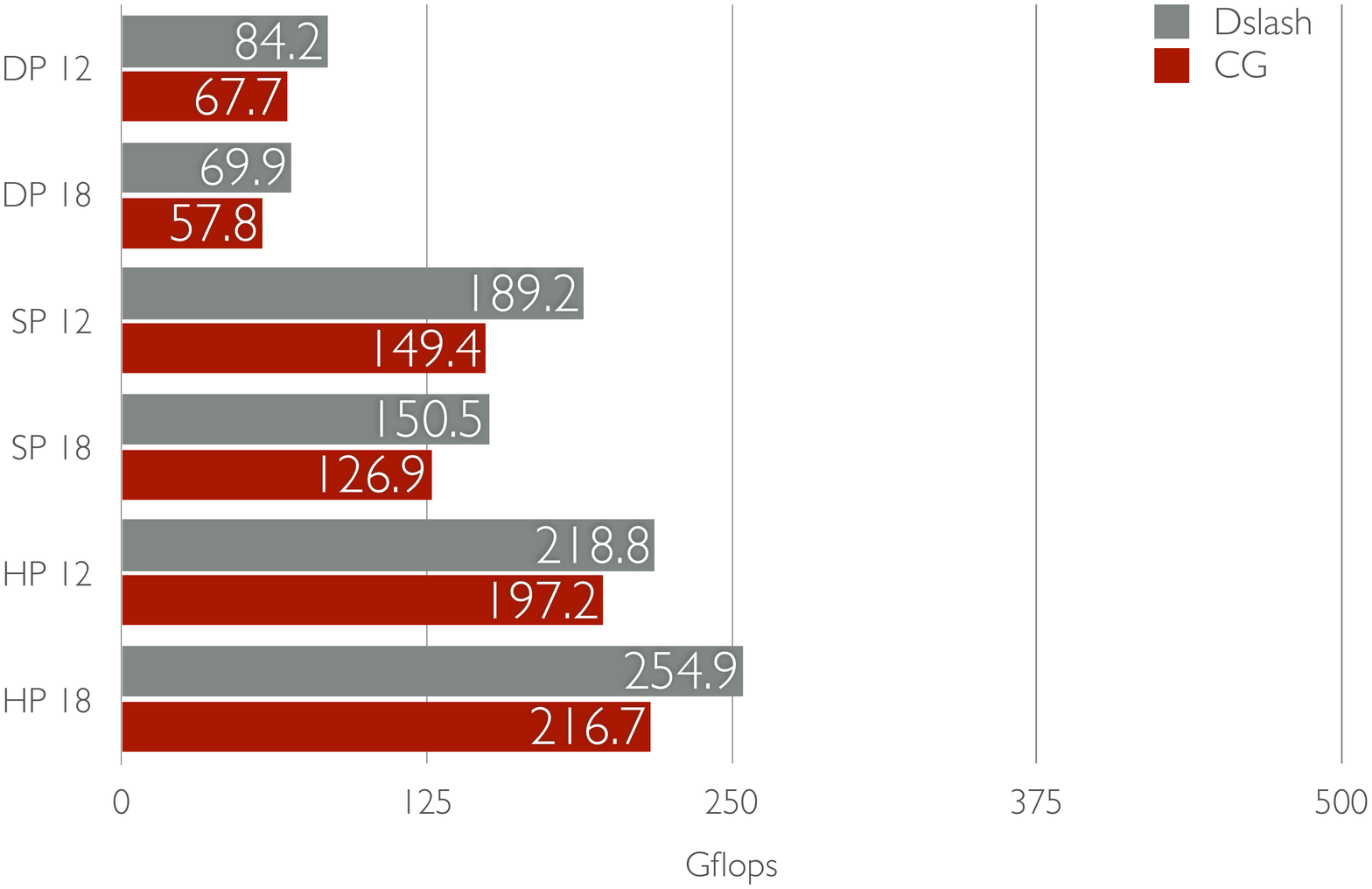}
	\caption{Twisted Mass $32^3\times 64$ Dual Socket Xeon Haswell E5-2630 2.4GHz AVX2.}
	\label{fig:tm_32x64_haswell}
\end{figure}

In \fig{fig:tm_32x64_haswell} we present an equivalent performance plot but on a Dual Socket Xeon Haswell CPU (E5-2630 at 2.4GHz) with kernels that we produced with the QPhiX code generator for AVX2 instructions. The performance gives around 60\% of the corresponding Xeon-Phi performance.
With the tmLQCD package \cite{Jansen:2009xp, Abdel-Rehim:2013wba}, we reach 28 Gflops in double and 38 Gflops in single precision, respectively, for the twisted mass Dslash operation. (Note that tmLQCD supports neither 12 compression nor half precision arithmetics).
Therefore our QPhiX kernels account for speedup factors of $3\times$ in DP, $4.9\times$ in SP and $6.7\times$ in HP (the latter in relation to the tmLQCD SP result), respectively.

\subsection{Weak scaling}
\begin{figure}[htb]
	\center
	\includegraphics[width=0.75\textwidth]{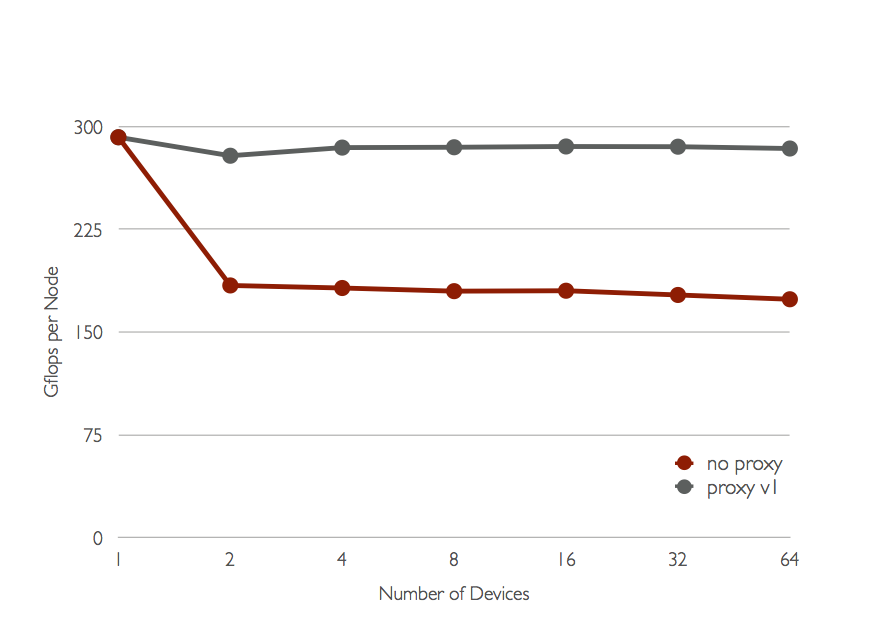}
	\caption{Weak Scaling Twisted Mass Dslash SP $48^3\times 96$ per device Xeon-Phi 7120P.}
	\label{fig:weak_mic}
\end{figure}

\begin{figure}[htb]
	\center
	\includegraphics[width=0.75\textwidth]{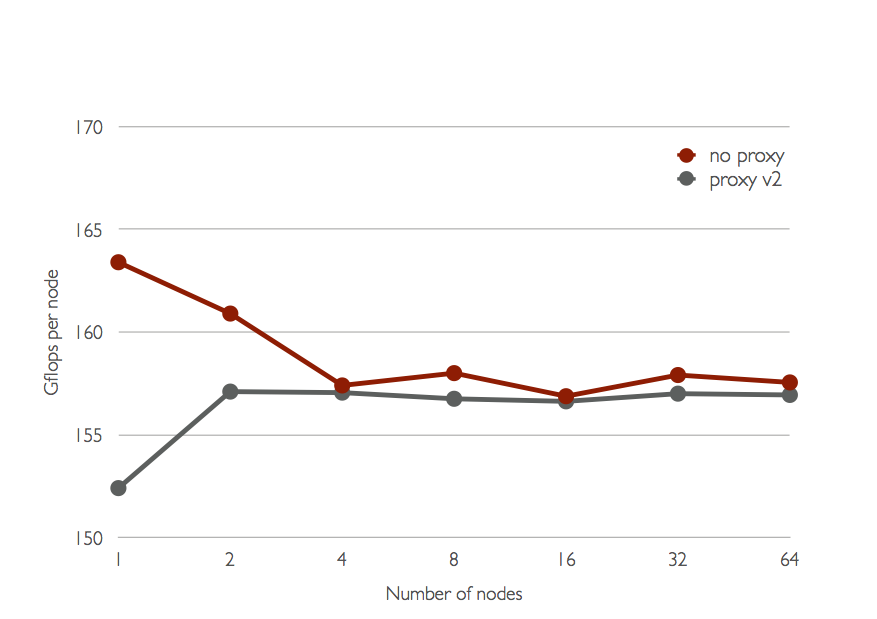}
	\caption{Weak Scaling Twisted Mass Dslash SP $32^3\times 64$ per node Dual Socket Xeon Haswell E5-2630 2.4GHz AVX2.}
	\label{fig:weak_haswell}
\end{figure}

\fig{fig:weak_mic} and \fig{fig:weak_haswell} show the weak scaling behaviour of the Dslash kernel on up to 64 Xeon-Phi 7120P devices 
and up to 64 Dual socket Haswell E5-2630 CPUs, respectively.
These tests have been performed on the ``Galileo'' cluster\footnote{\codefont{http://www.hpc.cineca.it/hardware/galileo}} at Cineca, Italy.
On both architectures we compare the performance to a run with a proxy \cite{joo:2013} that has been 
made available to us by Parallel Computing Lab, Intel Corporation. When running the code with the proxy, the latter will fully
occupy one of the compute cores. Loosing one of the 16 cores of the Dual Socket Xeon Haswell is too costly to profit from optimized MPI communication (\fig{fig:weak_haswell}).
On the Xeon-Phi (\fig{fig:weak_mic}), on the other hand, dedicating one of the 61 cores for communication pays off and weak scaling is better
than without proxy.
The kernel reaches 18.2 Tflops on 64 devices.

\subsection{Strong scaling}
\begin{figure}[htb]
	\center
	\includegraphics[width=0.75\textwidth]{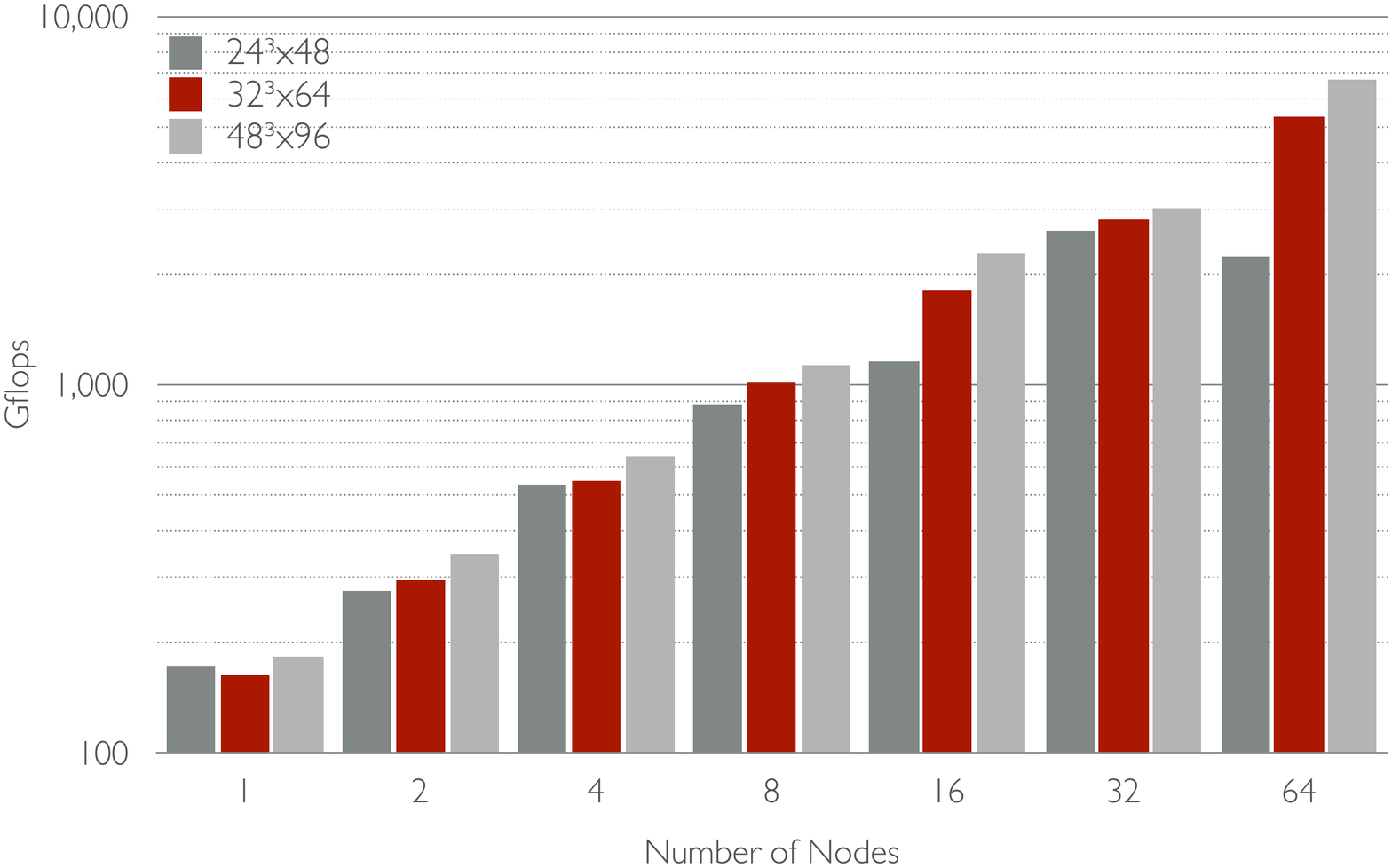}
	\caption{Strong Scaling Twisted Mass Dslash SP Compression12 Dual Socket Xeon Haswell E5-2630 2.4GHz AVX2.}
	\label{fig:strong_haswell_SP}
\end{figure}

\begin{figure}[htb]
	\center
	\includegraphics[width=0.75\textwidth]{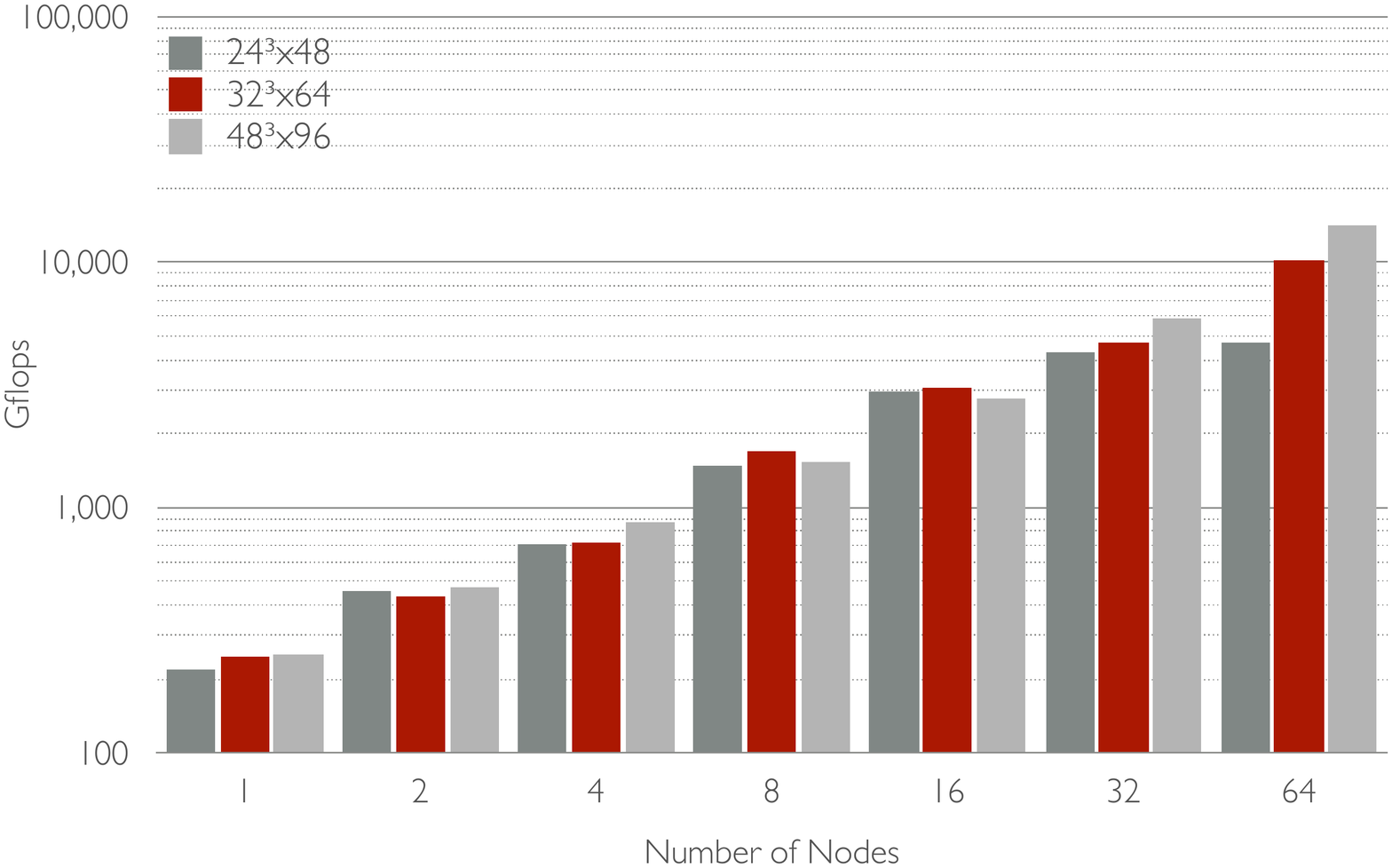}
	\caption{Strong Scaling Twisted Mass Dslash HP Compression18 Dual Socket Xeon Haswell E5-2630 2.4GHz AVX2.}
	\label{fig:strong_haswell_HP}
\end{figure}

In Figs.~\ref{fig:strong_haswell_SP} and \ref{fig:strong_haswell_HP} we present the strong scaling  behaviour of the Dslash kernel
on Dual Socket Xeon Haswell CPUs
in single (SP) and half precision (HP), respectively, for a range of lattice sizes ($24^3\times48$, $32^3\times64$ and $48^3\times96$).
While the strong scaling behaviour on the Xeon CPU is resonable, the Xeon-Phis require a too large local volume (compared to today's state of the
art lattice sizes) to reach high performance.
On the largest lattice we reach 6.8 (14.1) Tflops in single (half) precision on 64 nodes.

\section{Summary}
We have implemented \emph{Dslash} and \emph{AChiMBDPsi} kernels of the twisted mass fermion formulation for the QPhiX library.
The code passes the unit tests of the kernels, the full fermion matrix \eq{eq:Moo} and the Conjugate Gradient algorithm.
On the Xeon-Phi 7120P the Dslash kernel reaches 80\% of the theoretical peak bandwidth and on a Dual Socket Xeon Haswell CPU
with AVX2 generated code our QPhiX kernel outperforms the tmLQCD library by a factor of $4.9\times$ in single precision ($6.7\times$ 
when making use of half precision arithmetics).
Thanks to Intel's MPI proxy the weak scaling behaviour is good not only on the Xeon CPUs but also on Xeon-Phis.
While strong scaling is good on the Xeon Haswell CPUs on our test machine, the Xeon-Phis require too large local volumes for 
practical purposes at this stage.

\begin{acknowledgments}
We are very greatful to 
B\'{a}lint Jo\'{o}, 
Dhiraj D. Kalamkar
and
Karthikeyan Vaidyanathan
for sharing the QPhiX library and moreover
for helpful discussions and assistance.
Support by INFN and SUMA is acknowledged.

\end{acknowledgments}


\providecommand{\href}[2]{#2}\begingroup\raggedright\endgroup

\end{document}